\documentstyle[multicol,prb,aps,psfig]{revtex}

\begin{document}
\title{Spectral Functions for the Tomonaga--Luttinger and Luther--Emery
Liquids}

\author{Dror~Orgad}
\address
{Department of Physics, University of California,
Los Angeles, CA  90095-1547}

\date{\today}
\maketitle 

\begin{abstract}
We calculate the finite temperature single hole spectral function and the 
spin dynamic structure factor of spinfull one-dimensional Tomonaga-Luttinger 
liquid. Analytical expressions are obtained for a number of special cases. 
We also calculate the single hole spectral function of a spin gapped 
Luther-Emery liquid and obtain exact results at the free fermion point 
$K_s=1/2$. These results may be applied to the analysis of angle 
resolved photoemission and neutron scattering experiments on 
quasi-one-dimensional materials.
\end{abstract}

\pacs{PACS numbers: }

\begin{multicols}{2}

\section{Introduction}
The interacting one-dimensional electron gas (1DEG) has intrigued physicists 
since the pioneering work of Tomonaga\cite{Tomonaga50}. Through the 
collective effort of many of them, and especially due to the advancement  
of the bosonization technique\cite{Schotte69,Mattis74,Luther74}, a large 
number of its properties have been discovered. 

The past two decades have witnessed an experimental effort 
to identify and study, using various probes, quasi-one-dimensional 
systems. Among the materials studied one finds organic compounds 
such as polyacetylene\cite{Heeger88} and the Bechgaard 
salts\cite{Bourbonnais99} as well as several families of inorganic 
quasi-one-dimensional materials\cite{Monceau85,Gruner94} (for example, 
the transition metal bronzes). The edge states in the quantum Hall effect 
constitute yet another realization of one-dimensional interacting 
systems\cite{edges}. Progress made in nanoscale fabrication techniques 
have made it possible to artificially define one-dimensional channels 
in two-dimensional electronic heterostructures\cite{hetero}. 
Most recently, experimental evidence has emerged for the existence of stripe 
phases in doped antiferromagnets, such as the copper-oxide high temperature 
superconductors\cite{Emery99}. In these phases the doped holes segregate 
into quasi-one-dimensional metallic regions embedded in a predominantly 
antifferomagnetic background.  

In recent years angle resolved photoemission spectroscopy (ARPES) has 
matured into a powerful experimental method for probing the single 
particle properties of strongly correlated systems. ARPES measurements 
on the Bechgaard salts\cite{Zwick97}, the transition metalchalcogenides
\cite{TaSeI}, the blue bronzes\cite{blue} and the cuprates\cite{cuprates}
have provided evidence in favor of correlated one-dimensional physics 
although full agreement with the predicted theoretical picture is 
still lacking.  

The zero temperature correlation functions of the gapless 
Tomonaga-Luttinger liquid, and in particular the single hole spectral 
function which is measured by ARPES, have been calculated previously for the 
spinless\cite{Luther74} as well as for the spinfull 
case\cite{Meden92,Voit93}. Extending these results to finite temperatures 
is desirable on the following grounds. First, real experiments are 
carried out at finite temperatures and should be contrasted with theoretical 
predictions relevant for such conditions. Secondly, the Tomonaga-Luttinger 
liquid is a quantum critical system which consequently exhibits 
scaling behavior. It is a rare example where not only the scaling exponents 
but indeed the entire scaling functions can be computed explicitly. In 
Section II we obtain the scaling form of the single hole spectral function 
and the spin dynamic structure factor of the spinfull Tomonaga-Luttinger 
liquid. Closed form analytical expressions are derived for a number of special 
cases. These results complement previous numerical evaluations of the finite 
temperature spectral functions \cite{Schonhammer93,Nakamura97} and allow 
for easy investigation of their properties in few limits which are of 
physical interest. Integrals of these spectral functions are considered 
as well.      

When backward scattering of electrons becomes relevant the 1DEG develops 
a spin gap and is described by the Luther-Emery liquid\cite{Luther74b} 
(Umklapp processes may create a charge gap). While the spectrum of the 
Luther-Emery liquid can be readily derived at the special ``free fermion'' 
point $K_s=1/2$, the evaluation of electronic correlation functions there 
is non-trivial due to their highly non-local form in terms of the pseudo-
fermions. Several authors have made progress in this direction, see Refs. 
\ref{voitref},\ref{wiegmannref}, and in section III we extend and correct their 
study and obtain exact expressions for the single hole spectral function 
at the free fermion point for temperatures much smaller than the spin gap. 
Some technical details are relegated to appendices. 

\section{The Tomonaga--Luttinger Liquid}
\subsection{The model}
The Tomonaga--Luttinger model embodies the low energy and long wavelength 
physics of the 1DEG. It is composed of
two branches of left $(\eta=-1)$ and right $(\eta=+1)$ moving massless
Dirac fermions constructed around the left and right Fermi points of the 
1DEG. We will consider the case where the fermions carry spin $1/2$ and 
denote by $\sigma=\pm 1$ their spin polarization. In the absence of backward 
scattering and Umklapp processes the Hamiltonian density is given by 
$(\hbar=1)$
\begin{eqnarray}
\label{hamiltonian}
\nonumber
{\cal H}=&-&v_F\sum_{\eta,\sigma=\pm 1}\eta \psi_{\eta,\sigma}^{\dagger}i
\partial_x\psi_{\eta,\sigma} \\
\nonumber
&+&\frac{1}{2}\sum_{\eta=\pm 1}g_{2,c}\, \rho_{\eta}(x)\rho_{-\eta}(x)
+g_{4,c}\, \rho_{\eta}(x)\rho_{\eta}(x) \\
&+&2\sum_{\eta=\pm 1}g_{2,s}\,{\rm S}_{\eta}^z(x){\rm S}_{-\eta}^z(x)+g_{4,s}\,
{\rm S}_{\eta}^z(x){\rm S}_{\eta}^z(x) \; ,
\end{eqnarray}   
where $v_F$ is the noninteracting Fermi velocity and 
\begin{eqnarray}
\label{pdensity}
\rho_\eta&=&\sum_{\sigma}\psi_{\eta,\sigma}^{\dagger}\psi_{\eta,\sigma} 
\; , \\
\label{sdensity}
{\bf S}_{\eta}&=&\frac{1}{2}\sum_{\sigma,\sigma'}\psi_{\eta,\sigma}^{\dagger}
\mbox{\boldmath $\tau$}_{\sigma\sigma'}\psi_{\eta,\sigma'} \; ,
\end{eqnarray}
with \mbox{\boldmath $\tau$} being the Pauli matrices. Despite its 
appearance, the last term in (\ref{hamiltonian}) does not break the $SU(2)$ 
spin symmetry, since ${{\rm S}_{\eta}^z}^2(x)=\lim_{x'\rightarrow x}{\bf S}_
{\eta}(x)\cdot{\bf S}_{\eta}(x')/2+\rho_{\eta}^2(x)/8$; the $g_{2,s}$ term 
is the only term which breaks this symmetry.

In order to compute the single particle properties of the model one uses the 
bosonization identity\cite{Emery79,Delft98}
\begin{equation}
\label{bidentity}
\psi_{\eta,\sigma}=\frac{1}{\sqrt{2\pi a}}F_{\eta,\sigma}\exp[i\eta k_F x
-i\Phi_{\eta,\sigma}(x)] \; ,
\end{equation}
which expresses the fermionic fields in terms of self-dual fields 
$\Phi_{\eta,\sigma}(x)$ obeying $[\Phi_{\eta,\sigma}(x),\Phi_{\eta',\
\sigma'}(x')]=-i\pi\delta_{\eta,\eta'}\delta_{\sigma,\sigma'}{\rm sign}(x-x')$.
The Klein factors $F_{\eta,\sigma}$ are responsible for reproducing the 
correct anti-commutation relations between different fermionic species and 
$a$ is a short distance cutoff that is taken to zero at the end of the 
calculation. Here and throughout the paper we consider the limit where the 
size of the system $L$ is taken to infinity and correspondingly ignore terms 
of the order $1/L$ (for a discussion of finite size effects in the finite 
temperature case see Refs. \ref{Mattsson97ref}-\ref{Eggert97ref}). 

The demonstration of the celebrated charge-spin separation in the model is 
facilitated by expressing $\Phi_{\eta,\sigma}$ in terms of the bosonic 
fields $\phi_c\, ,\phi_s$ and their conjugated momenta $\partial_x\theta_c
\, ,\partial_x\theta_s$
\begin{equation}
\label{decomp}
\Phi_{\eta,\sigma}=\sqrt{\pi/2}\,[(\theta_c-\eta\phi_c)+\sigma(\theta_s-\eta
\phi_s)] \; ,
\end{equation}
in terms of which the charge and spin densities are given by
\begin{eqnarray}
\label{cden}
\rho(x)&=&\sum_{\eta}\rho_{\eta}(x)=\sqrt{2/\pi}\,\partial_x\phi_c \; , \\
\label{sden}
{\rm S}^z(x)&=&\sum_{\eta}{\rm S}^z_{\eta}(x)=\sqrt{1/2\pi}\,
\partial_x\phi_s \; .
\end{eqnarray}
The decomposition (\ref{decomp}) also facilitates the diagonalization of 
the Hamiltonian. It becomes a sum of two independent pieces describing 
noninteracting charge and spin density waves which are the elementary 
excitations of the system 
\begin{equation}
\label{diagh}
{\cal H}=\sum_{\alpha=c,s}\frac{v_\alpha}{2}\left[K_\alpha(\partial_x
\theta_\alpha)^2+\frac{(\partial_x\phi_\alpha)^2}{K_\alpha}\right] \; .
\end{equation}
The velocities of the collective modes are
\begin{equation}
\label{vel}
v_\alpha=\sqrt{\left(v_F+\frac{g_{4,\alpha}}{\pi}\right)^2-\left(\frac
{g_{2,\alpha}}{\pi}\right)^2} \; ,
\end{equation}
and the parameters $K_\alpha$, which determine the power-law behavior of the 
correlation functions, read
\begin{equation}
\label{k}
K_\alpha=\sqrt{\frac{\pi v_F+g_{4,\alpha}-g_{2,\alpha}}
{\pi v_F+g_{4,\alpha}+g_{2,\alpha}}} \; .
\end{equation} 
The effective parameters that enter the Hamiltonian depend
on the specifics of the model for which (\ref{hamiltonian}) is 
a low energy fixed point. We note, however, that spin-rotation invariance 
dictates $g_{2,s}=0$ and consequently $K_s=1$. 

\subsection{The space-time correlation functions}
The bosonized expression for the fermion field operators (\ref{bidentity})
and the fact that the theory reduces to that of free bosons make it 
straightforward to derive expressions for the space-time 
response functions (which we denote with a tilde). We will focus on the 
finite temperature single hole 
Green function 
\begin{equation}
\label{rgfun}
\tilde G^<_{\eta}(x,t;T)=\langle\psi_{\eta,\sigma}^{\dagger}(x,t)
\psi_{\eta,\sigma}(0,0)\rangle \; .
\end{equation}
We compute this Green function, rather than the more usual time ordered 
or retarded Green functions, because it includes only the one-hole states. 
This fact makes it relevant to ARPES whose
cross section is directly proportional to the Fourier transform of 
$\tilde G^<$. Other Green functions can be easily obtained from it, 
since for the model (\ref{hamiltonian}) $\tilde G^>_{\eta}(x,t;T)=
\langle\psi_{\eta,\sigma}(0,0)\psi_{\eta,\sigma}^{\dagger}(x,t)\rangle=
\tilde G^<_{\eta}(-x,-t;T)$.

We will also consider the $2k_F$ component of the transverse spin dynamic 
structure factor, which is measured by polarized neutron scattering
\begin{eqnarray}
\label{sxy}
\nonumber
\tilde{\cal S}(x,t;T)&=&\langle {{\rm S}^x_{2k_F}}^{\dagger}
(x,t) {\rm S}^x_{2k_F}(0,0)\rangle \\
&+&\langle {{\rm S}^y_{2k_F}}^{\dagger}(x,t)
{\rm S}^y_{2k_F}(0,0)\rangle  \; .  
\end{eqnarray}
where 
\begin{equation}
\label{s2kfdef}
{\bf S}_{2k_F}=\frac{1}{2}\sum_{\sigma,\sigma'}\psi_{1,\sigma}^{\dagger}
\mbox{\boldmath $\tau$}_{\sigma\sigma'}\psi_{-1,\sigma'} \; ,
\end{equation}
and the $\mbox{\boldmath $\tau$}$ are the Pauli matrices. 

The Tomonaga-Luttinger liquid (\ref{hamiltonian}) is a quantum critical system.
It also exhibits spin-charge separation. This implies a scaling form for the 
response functions with separate spin and charge pieces. Specifically one 
finds\cite{Emery79}
\begin{eqnarray}
\label{scalegr}
\nonumber
\tilde G^<_{\eta}(x,t;T)=&&\frac{1}{2\pi a}e^{-i\eta k_F x}\left(\frac
{a}{\lambda_{T,c}}\right)^{2\gamma_c+\frac{1}{2}}\left(\frac{a}{\lambda_{T,s}}
\right)^{2\gamma_s+\frac{1}{2}} \\
&&\times\tilde g_c\left(\frac{x}{\lambda_{T,c}},\frac{v_c t}
{\lambda_{T,c}}\right)\tilde g_s\left(\frac{x}{\lambda_{T,s}},\frac{v_s t}
{\lambda_{T,s}}\right) \, ,
\end{eqnarray}
where we introduced the thermal lengths ($k_B=1$) 
\begin{equation}
\label{tl}
\lambda_{T,\alpha}=\frac{v_\alpha}{\pi T} \; ,
\end{equation}
and the exponents
\begin{equation}
\label{gammadef}
\gamma_\alpha=\frac{1}{8}(K_\alpha+K_\alpha^{-1}-2) \;,
\end{equation}
defined so that $\gamma_\alpha=0$ for noninteracting fermions. Since 
the spin and charge sectors are formally invariant under separate Lorentz 
transformations the functions $\tilde g_\alpha$ also spilt into right 
and left moving parts 
\begin{equation}
\label{gxt}
\tilde g_\alpha(x,t)=\tilde h_{\gamma_\alpha+\frac{1}{2}}(\eta x-t) \,
\tilde h^{*}_{\gamma_\alpha}(\eta x+t) \; ,
\end{equation}
where
\begin{equation}
\label{hdef}
\tilde h_\gamma(x)=\left[-i\sinh(x+ia)\right]^{-\gamma} \; .
\end{equation}
Similarly, for the spin correlation function one finds\cite{Luther74,Emery79}
\begin{eqnarray}
\label{scalesr}
\nonumber
\tilde {\cal S}(x,t;T)=&&\frac{1}{(2\pi a)^2}e^{-2i k_F x}
\left(\frac{a}{\lambda_{T,c}}\right)^{2\beta_c}\left(\frac{a}
{\lambda_{T,s}}\right)^{2\beta_s} \\
\nonumber
&&\times\tilde C_c\left(\frac{x}{\lambda_{T,c}},\frac{v_c t}
{\lambda_{T,c}}\right)\tilde C_s\left(\frac{x}{\lambda_{T,s}},
\frac{v_s t}{\lambda_{T,s}}\right) \, , \\
\end{eqnarray}
where
\begin{equation}
\label{cxt}
\tilde C_\alpha(x,t)=\tilde h_{\beta_\alpha}(x-t) \,\tilde h^{*}
_{\beta_\alpha}(x+t) \; .
\end{equation}
Here we introduced the exponents 
\begin{equation}
\label{betadef}
\beta_c=\frac{K_c}{2} \;\;\;\; , \;\;\;\; \beta_s=\frac{1}{2K_s} \; .
\end{equation}

We note that the perpendicular component of the spin dynamic structure factor 
\begin{equation}
\label{szz}
\tilde{\cal S}^{z}(x,t;T)=\langle {{\rm S}^z_{2k_F}}^{\dagger}(x,t) 
{\rm S}^z_{2k_F}(0,0)\rangle \; ,
\end{equation}
is obtained from the result for the transverse part (\ref{scalesr}) after 
multiplying it by an overall factor of $1/2$ and using the exponents 
\begin{equation}
\label{betazdef}
\beta_c=\frac{K_c}{2} \;\;\;\; , \;\;\;\; \beta_s=\frac{K_s}{2} \; ,
\end{equation}
instead of (\ref{betadef}). This holds true also for the spectral 
functions calculated in the following subsections. Of coarse, 
in the spin rotation invariant case $\tilde{\cal S}^z=\tilde{\cal S}/2$.

\subsection{The spectral functions}
It is a general feature of the field-theoretic approach to this problem 
that relatively simple expressions are obtained for the space-time 
dependent correlation functions. For comparison with experiments, however,
we are typically interested in the Fourier transform of these correlation 
functions. Conceptually, this is simple, and indeed, to evaluate the 
Fourier transform, we need only to perform a two dimensional integral.
However, it is not generally simple to carry out this calculation 
analytically. Below we consider the cases where this can be done.

As we noted before, the critical nature of the model leads to a scaling form 
for the spectral functions. A simplifying feature introduced by the 
spin-charge separation is the ability to express these scaling functions as 
a convolution of spin and charge parts 
\begin{eqnarray}
\label{scalegk}
\nonumber
G^<(\tilde k,\tilde\omega;T)&=&\int_{-\infty}^{\infty}dx\,dt\, 
e^{i[\eta(k+k_F)x-\omega t]}\tilde G^<_{\eta}(x,t;T) \\
\nonumber
&=&\frac{a}{(2\pi)^3 v_c}\left(\frac{a}{\lambda_{T,c}}\right)^{2\gamma_c-
\frac{1}{2}}\left(\frac{a}{\lambda_{T,s}}\right)^{2\gamma_s-\frac{1}{2}} \\
\nonumber
&\times&\int_{-\infty}^{\infty} dq\, d\nu\, g_c(q,\nu)\,g_s
(\tilde k-rq,\tilde\omega-\nu) \\
\end{eqnarray}
where we introduce the velocity ratio $r=v_s/v_c$, define the scaling 
variables
\begin{equation}
\label{scalevar}
\tilde k=\frac{v_s k}{\pi T} \;\;\;\; , \;\;\;\; 
\tilde\omega=\frac{\omega}{\pi T} \; ,
\end{equation}
and  
\begin{equation}
\label{gk}
g_\alpha(k,\omega)=\frac{1}{2}h_{\gamma_\alpha+\frac{1}{2}}
\left(\frac{\omega+k}
{2}\right)h_{\gamma_\alpha}\left(\frac{\omega-k}{2}\right) \; .
\end{equation}
$h_{\gamma}(k)$, the Fourier transform of $\tilde h_{\gamma}(x)$,
is evaluated in Appendix A, where some of its properties are listed as well. 
For non-integer values of $\gamma$ it is given by 
\begin{equation}
h_\gamma(k)={\rm Re}\left[(2i)^{\gamma}B\left(\frac{\gamma-ik}{2},1-\gamma
\right)\right] \; ,
\end{equation}
where $B(x,y)$ is the Beta function.

In a similar fashion we obtain for the spin susceptibility
\begin{eqnarray}
\label{scalesk}
\nonumber
{\cal S}(\tilde k,\tilde\omega;T)&=&\int_{-\infty}^{\infty}dx\,dt\, 
e^{i[(k+2 k_F)x-\omega t]}\tilde {\cal S}(x,t;T) \\
\nonumber
&=&\frac{1}{(2\pi)^4 v_c}\left(\frac{a}{\lambda_{T,c}}\right)^
{2\beta_c-1}\left(\frac{a}{\lambda_{T,s}}\right)^{2\beta_s-1} \\
\nonumber
&\times& \int_{-\infty}^{\infty} dq\, d\nu\, C_c(q,\nu)\,C_s
(\tilde k-rq,\tilde\omega-\nu) \\
\end{eqnarray}
where 
\begin{equation}
\label{Ck}
C_\alpha(k,\omega)=\frac{1}{2}h_{\beta_\alpha}\left(\frac
{\omega+k}{2}\right)h_{\beta\alpha}\left(\frac{\omega-k}{2}\right) \; .
\end{equation}
Henceforth $k$ is measured relative to $k_F$ and $2k_F$ when 
computing $G^<$ and ${\cal S}$, respectively. We also define the 
Fourier transform with respect to $\eta x$. This has the effect of making 
$k$ positive outside the Fermi surface and negative inside; thus giving 
the same expression for the spectral function of left and right moving holes. 

Here we would like to note the non-commutativity of the limits 
$k,\omega\rightarrow 0$ and $T\rightarrow 0$ 
in calculating the zero temperature dc response of the system. 
Due to the scaling form of the spectral functions taking the 
zero temperature limit first gives a result which is determined 
by the $\tilde k,\tilde\omega\rightarrow\infty$ behavior of the scaling 
functions, while it is the $\tilde k,\tilde\omega\rightarrow 0$ behavior 
which is relevant in case $k$ or $\omega$ is set to zero from the outset. 

Further analytic progress in calculating the spectral functions can be 
achieved in the following cases:

\subsubsection{The case $v_c=v_s$}
Since there is no symmetry between the charge and spin sectors, we generally 
expect that the spin and charge velocities are different. This greatly 
complicates the explicit calculation of the Fourier transforms, as the model 
is not ``Lorentz invariant'' under transformations that involve both the spin 
and charge sectors. Such an invariance is restored if $v_c=v_s=v$. In this 
case the correlation functions have the same form as those of a related 
model of spinless electrons, for which the zero temperature spectral 
functions have been calculated by Luther and Peschel \cite{Luther74}.
    
The major simplification that follows directly from the fact that the charge 
and spin velocities are equal is that the two-dimensional Fourier transform 
reduces to a product of two one-dimensional transforms - one for the right 
moving piece and one for the left moving piece. Specifically one finds, 
with $\gamma_G=\gamma_c+\gamma_s$, that
\begin{eqnarray}
\label{geqv}
\nonumber
G^<(\tilde k,\tilde\omega;T)&=&\frac{1}{4\pi^2 T}\left(\frac{a}{\lambda_T}
\right)^{2\gamma_G} \\
\nonumber
&\times& h_{\gamma_G+1}\left(\frac{\tilde\omega+\tilde k}{2}\right)
h_{\gamma_G}\left(\frac{\tilde\omega-\tilde k}{2}\right) \; . \\
\end{eqnarray}

A similar analytic form can be obtained for the spin correlation function
\begin{eqnarray}
\label{seqv}
\nonumber
{\cal S}(\tilde k,\tilde\omega;T)&=&\frac{1}{8\pi^2 v}
\left(\frac{a}{\lambda_T}\right)^{2(\beta_{\cal S}-1)} \\ 
&\times&h_{\beta_{\cal S}}\left(\frac{\tilde\omega+\tilde k}{2}
\right)h_{\beta_{\cal S}}\left(\frac{\tilde\omega-\tilde k}{2}
\right) \; , 
\end{eqnarray}
where $\beta_{\cal S}=\beta_c+\beta_s$.

\subsubsection{The spin-rotationally invariant case $(K_s=1)$}
We already noted that when the system is invariant under spin rotations 
$K_s=1$ ($\gamma_s=0$). At this important special point there is no mixing 
between left and right moving spin excitations. As a result the expression 
for the hole spectral function is simplified, due to the appearance of the 
factor $h_0$ in the spin part of (\ref{gk}), and can be expressed as a single 
integral
\begin{eqnarray}
\label{gkgamma0}
\nonumber
&&G^<(\tilde k,\tilde\omega;T)=
\frac{r^{1/2}}{4\pi^3 T}\left(
\frac{a}{\lambda_{T,c}}\right)^{2\gamma_c}\int_{-\infty}^{\infty}dq\,
h_{\frac{1}{2}}\left(\tilde k-2 r q\right) \\
\nonumber
&&\hspace{0.55 cm}\times h_{\gamma_c+\frac{1}{2}}
\left[\frac{\tilde\omega-\tilde k}{2}+(1+r)q\right]  
h_{\gamma_c}\left[\frac{\tilde\omega-\tilde k}{2}
-(1-r)q\right] \; . \\
\end{eqnarray}   

When $K_c=1$ as well, the integral in (\ref{gkgamma0}) is trivial and 
we obtain
\begin{equation}
\label{gk00}
G^<(\tilde k,\tilde\omega;T)=\frac{1}{2\pi^2 T}\frac{r^{1/2}}{|1-r|}
h_{\frac{1}{2}}\left(\frac{\tilde\omega-\tilde k}{1-r}\right)
h_{\frac{1}{2}}\left(\frac{r\tilde\omega-\tilde k}{r-1}\right) \; .
\end{equation}

\begin{figure}
\narrowtext
\setlength{\unitlength}{1in}
\begin{picture}(3.2,4.2)(0,-1.2)
\put(-0.15,-1){\psfig{figure=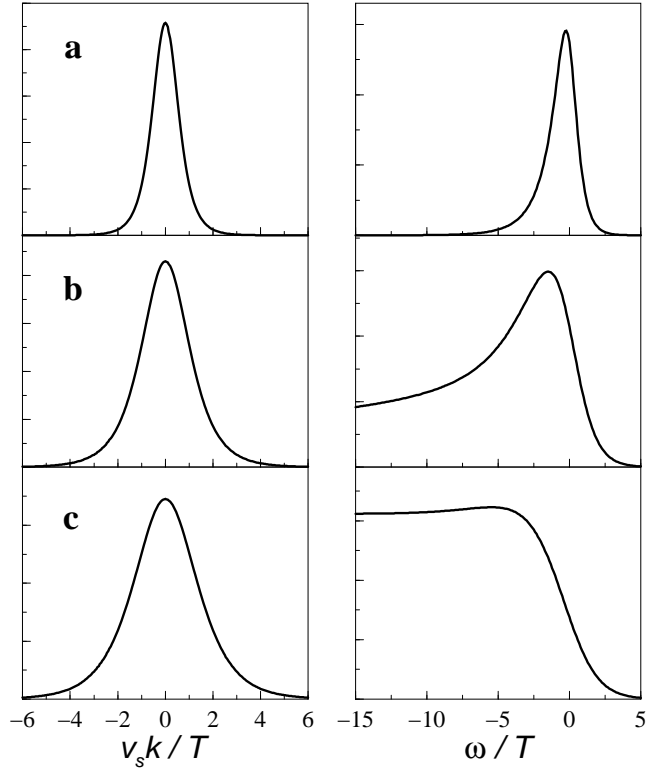,width=3.4in}}
\end{picture}
\caption
{MDCs at $\omega=0$ (left) and EDCs at $k=0$ (right), for a spin 
rotationally invariant Tomonaga-Luttinger liquid,   
plotted versus $v_s k/T$ and 
$\omega/T$ respectively, with $v_c/v_s=3$ and a) $\gamma_c=0$, 
b) $\gamma_c=0.25$, and c) $\gamma_c=0.5$.}
\label{fig:1}
\end{figure}

The case $\gamma_s=\gamma_c=0$ is unique in the sense that there is no 
mixing between left and right moving excitations. As a result there are 
severe kinematic constraints on $G^<$ that make it non-vanishing, in the 
zero temperature limit, only in a wedge in $k-\omega$ plane defined by 
the lines $\omega=v_s k$ and $\omega=v_c k$ for $k<0$. 
While $K_s=1$ reflects a symmetry 
of the problem it is unlikely that $K_c=1$ ($\gamma_c=0$) is realized in 
any interacting system. However, if the effective interactions are not too 
strong then $\gamma_c$ may be small. For instance, for the Hubbard 
model\cite{Kawakami90}, even in the $U/t\rightarrow\infty$ limit, 
$\gamma_c=1/16$. For such systems we expect (\ref{gk00}) to be qualitatively 
correct with the exception of the behavior of $G^<$ outside 
the above mentioned wedge. In contrast to the result for $\gamma_c=0$ 
the $T=0$ support of $G^<$, for non-zero $\gamma_c$, extends up to the line 
$\omega=-v_c k$ with $k>0$. Nevertheless, if the mixing between left and 
right moving charge excitations (i.e $\gamma_c$) is small the amplitude of 
$G^<$ outside the wedge is small and its gross features resemble 
the $\gamma_c=0$ result.
 
The customary way to present ARPES data is by plotting momentum distribution 
curves (MDCs) and energy distribution curves (EDCs). These curves are cuts 
in $G^<(k,\omega)$ of constant $\omega$ and constant $k$ respectively. 
In Fig. 1 we present MDCs at the Fermi energy ($\omega=0$) and EDCs at 
the Fermi wave-vector $(k=0)$ for a spin rotationally invariant 
Tomonaga-Luttinger liquid for various values of the parameter $\gamma_c$.

\subsubsection{The case $v_s/v_c\rightarrow 0$}
In the limit where one of the velocities is much smaller than the other
the calculation becomes again more tractable. In many physical systems 
$v_c>v_s$ and we will consider this case (for example for the $t-J$ model 
away from half filling we have $v_s/v_c\sim J/t<1$). In the limit $r=v_s/v_c
\rightarrow 0$ the spin piece in Eq. (\ref{scalegk}) becomes 
$q$-independent and any singularities in $G^<$ disperse with the slow
velocity $v_s$. The $q$ integral is then readily evaluated by expressing 
the factors in $g_c$ as Fourier transforms of their real-space counterparts. 
The result is 
\begin{eqnarray}
\label{gkr0}
\nonumber
&&G^<(\tilde k,\tilde\omega;T)=\frac{a}{(2\pi)^2 v_c}
\left(\frac{a}{\lambda_{T,c}}\right)^{2\gamma_c-\frac{1}{2}}
\left(\frac{a}{\lambda_{T,s}}\right)^{2\gamma_s-\frac{1}{2}} \\
\nonumber
&&\hspace{0.7 cm}\times\int_{-\infty}^{\infty}d\nu\, h_{2\gamma_c+\frac{1}{2}}
(\tilde\omega-\tilde k-2\nu)\, h_{\gamma_s+\frac{1}{2}}(\tilde k+\nu)\,
h_{\gamma_s}(\nu) \; . \\
\end{eqnarray}
A similar calculation gives for the spin spectral function
\begin{eqnarray}
\label{skr0}
\nonumber
&&{\cal S}(\tilde k,\tilde\omega;T)=
\frac{1}{(2\pi)^3 v_c}\left(\frac{a}{\lambda_{T,c}}
\right)^{2\beta_c-1}\left(\frac{a}{\lambda_{T,s}}\right)^{2\beta_s-1} \\
&&\hspace{0.6 cm}\times\int_{-\infty}^{\infty}d\nu\, h_{2\beta_c}
(\tilde\omega-\tilde k-2\nu)\, 
h_{\beta_s}(\tilde k+\nu)\, h_{\beta_s}(\nu) \; .
\end{eqnarray}

At the spin-rotationally invariant point $(K_s=1)$ the integral in 
(\ref{gkr0}) can be performed to give a result for $G^<$ in the limit 
$v_s/v_c\rightarrow 0$ but arbitrary $\gamma_c$
\begin{equation}
\label{gkr0gs0}
\nonumber
G^<(\tilde k,\tilde\omega;T)=
\frac{r^{1/2}}{2\pi^2 T}\left(\frac{a}{\lambda_{T,c}}\right)^{2\gamma_c}
h_{2\gamma_c+\frac{1}{2}}(\tilde\omega-\tilde k)\,
h_{\frac{1}{2}}(\tilde k) \; .
\end{equation}

\subsection{Integrals of the spectral functions}
Integrals of the spectral functions, such as the density of states and the 
momentum occupation number, are of interest too. These simpler quantities 
give us a qualitative view of the spectrum, without too many complicated 
details.

\subsubsection{$\rho^<(\omega)$ and ${\cal S}_0(\omega)$}
The calculation of the density of states $\rho^<(\omega)$ and its analogous 
quantity for the spin susceptibilities  ${\cal S}_0(\omega)$ is 
straightforward. The results are
\begin{eqnarray}
\label{rho}
\nonumber
&&\rho^<(\tilde\omega;T)=\int_{-\infty}^{\infty}\frac{dt}{2\pi}e^{i\omega t}
\tilde G^<_\eta(0,t)=\int_{-\infty}^{\infty}dk\, G^<(k,-\omega) \\
\nonumber
&&\hspace{1 cm}=\frac{1}{4\pi^2}\frac{1}{\sqrt{v_c v_s}}
\left(\frac{a}{\lambda_{T,c}}
\right)^{2\gamma_c}\!\left(\frac{a}{\lambda_{T,s}}\right)^{2\gamma_s}\! 
h_{2\gamma_G+1}(-\tilde\omega) \; ,\\
\end{eqnarray}
\begin{eqnarray}
\label{s0}
\nonumber
&&{\cal S}_0(\tilde\omega;T)=\int_{-\infty}^{\infty}\frac{dt}{2\pi}e^
{i\omega t}\tilde {\cal S}(0,t)=\int_{-\infty}^{\infty}dk\, 
{\cal S}(k,-\omega) \\
\nonumber
&&\hspace{0.5cm}=\frac{1}{8\pi^3 a}\frac{1}{\sqrt{v_c v_s}}
\left(\frac{a}{\lambda_{T,c}}
\right)^{2\beta_c-\frac{1}{2}}\!\left(\frac{a}{\lambda_{T,s}}\right)^
{2\beta_s-\frac{1}{2}}\!
h_{2\beta_{\cal S}}(-\tilde\omega)\; . \\
\end{eqnarray}

\subsubsection{$n(k)$ and $S(k)$ for the case $v_c=v_s$}
The evaluation of the momentum occupation number $n(k)$ and the spin 
structure factor $S(k)$ is complicated by the fact that they can not 
be expressed in terms of the functions $h_{\gamma}(k)$. Contrary to the 
scaling functions considered above the scaling functions of these quantities 
depend on the cut-off $a^{-1}$. It is, however, possible to compute them
in the case where the charge and spin velocities are equal.
\begin{eqnarray}
\label{nk1}
\nonumber
n(\tilde k;T)&=&\int_{-\infty}^{\infty}\!\!dx\,e^{i\eta(k+k_F)x}
\tilde G^<_\eta(x,0)
=\int_{-\infty}^{\infty}\frac{d\omega}{2\pi}G^<(k,\omega) \\
\nonumber
&=&-\frac{2^{\gamma_G+1}}{\pi}\left(\frac{a}{\lambda_T}\right)
^{2\gamma_G} \\
\nonumber
&\times&\!{\rm Im}\left\{\int_0^{\infty}\! 
dx \frac{e^{i\tilde k x}\sinh(
x-ia/\lambda_T)}{[\cosh(2x)-\cosh(2ia/\lambda_T)]^{\gamma_G+1}} 
\!\right\} \; ,\\
\end{eqnarray} 
where again $\gamma_G=\gamma_c+\gamma_s$.
Since the integrand in (\ref{nk1}) is imaginary in the interval 
$[ia/\lambda_T,0]$ of the imaginary axis, we can add the integral along 
this interval to the one already present in (\ref{nk1}) without 
affecting $n(k)$. 
\begin{eqnarray}
\label{nk2}
\nonumber
n&&(\tilde k;T)=\frac{2^{\gamma_G-1}}{\pi}\left(\frac{a}{\lambda_T}\right)
^{2\gamma_G} \\
\nonumber
&&\times{\rm Im}\left\{\!\int_{2ia/\lambda_T}^{\infty}\!\!\!\!\!
dx \frac{e^{\left(i\frac{\tilde k}{2}-\frac{1}{2}\right) x+ia/
\lambda_T}-
e^{\left(i\frac{\tilde k}{2}+\frac{1}{2}\right) x-ia/\lambda_T}}
{[\cosh(x)-\cosh(2ia/\lambda_T)]^{\gamma_G+1}}
\!\right\} \; . \\
\end{eqnarray}
Formally, for $\gamma_G<0$, Eq. (\ref{nk2}) is related to the 
integral representation of the associated Legendre function (defined with 
a branch cut from $-\infty$ to 1)
$Q_{\nu}^{\mu}(z)$, see Ref. \ref{erdref}. Since $n(k)$ is regular in 
$\gamma_G$ we can use analytic continuation to obtain for all 
non-integer $\gamma_G$  
\begin{eqnarray}
\label{nk3}
\nonumber
n(\tilde k;T)&=&\frac{2^{\gamma_G-1/2}}{\pi^{3/2}}\Gamma(-\gamma_G)
\left(\frac{a}{\lambda_T}\right)^{2\gamma_G} \\
\nonumber
&\times&{\rm Re}\Biggl\{ e^{-i\pi\gamma_G}
[\sinh(2ia/\lambda_T)]^{-(\gamma_G+1/2)}  \\
\nonumber
&&\hspace{0.25in}\times\Bigl[ e^{-ia/\lambda_T}Q_{-i\tilde k/2-1}
^{\gamma_G+1/2}\left[\cos(2a/\lambda_T)\right] \\
&&\hspace{0.32in}-e^{ia/\lambda_T}Q_{-i\tilde k/2}
^{\gamma_G+1/2}\left[\cos(2a/\lambda_T)\right]\Bigr]\Biggr\} \; .
\end{eqnarray}
It can be shown that for ${\gamma_G}\lesssim 1/2$ 
and $k\lesssim(\lambda_T a)^{-1/2}$ this expression reduces, 
in the limit $a\rightarrow 0$, to
\begin{eqnarray}
\label{nk4}
\nonumber
n(\tilde k;T)&=&\frac{\Gamma(-\gamma_G)}{2\pi^{3/2}}{\rm Re}\Biggl\{\Gamma
\left(\gamma_G+1/2\right)e^{-i\pi(\gamma_G+1/2)} \\ 
\nonumber
&+&\!i\left(\frac{a}{\lambda_T}\right)^{2\gamma_G}\!\frac{\Gamma(1/2-\gamma_G)
\Gamma[(1+2\gamma_G-i\tilde k)/2]}{\Gamma[(1-2\gamma_G-i\tilde k)/2]}
\Biggr\} \; . \\
\end{eqnarray}
Using similar manipulations we obtain, for non-integer $\beta_{\cal S}=
\beta_c+\beta_s$, the spin structure factor
\begin{eqnarray}
\label{sk1}
\nonumber
S(\tilde k;T)&=&\int_{-\infty}^{\infty}\!\!dx\,e^{i(k+2k_F)x}\tilde S(x,0)
=\int_{-\infty}^{\infty}\frac{d\omega}{2\pi}S(k,\omega) \\
\nonumber
&=&\frac{2^{\beta_{\cal S}-3/2}}{\pi^{5/2}a}\Gamma(1-\beta_{\cal S})
\left(\frac{a}{\lambda_T}\right)^{2\beta_{\cal S}-1} \\
\nonumber
&\times&{\rm Re}\Biggl\{e^{i\pi(1/2-\beta_{\cal S})}[\sinh(2ia/\lambda_T)]
^{1/2-\beta_{\cal S}} \\
&&\hspace{0.3in} \times Q_{-i\tilde k/2-1/2}^{\beta_{\cal S}-1/2}
\left[\cos(2a/\lambda_T)\right]\Biggr\} \; ,
\end{eqnarray}
which for $a \rightarrow 0$, $\beta_{\cal S}\lesssim 3/2$ and $k\lesssim(
\lambda_T a)^{-1/2}$ tends to
\begin{eqnarray}
\label{sk2}
\nonumber
S(\tilde k;T)&=&\frac{\Gamma(1-\beta_{\cal S})}{4\pi^{3/2}a}{\rm Re}\Biggl\{
\Gamma\left(\beta_{\cal S}-1/2\right)e^{i\pi(1/2-\beta_{\cal S})} \\ 
\nonumber
&+&\left(\frac{a}{\lambda_T}\right)^{2\beta_{\cal S}-1}\frac{\Gamma(1/2-
\beta_{\cal S})\Gamma(\beta_{\cal S}-i\tilde k/2)}{\Gamma(1-
\beta_{\cal S}-i\tilde k/2)}\Biggr\} \; . \\
\end{eqnarray}

\section{The Luther--Emery Liquid}

\subsection{The model}
In the Tomonaga-Luttinger liquid both the charge and spin excitations 
are gapless. Spin and charge gaps may open up as a result of adding to 
the Tomonaga-Luttinger Hamiltonian (\ref{hamiltonian}) terms describing 
backward and Umklapp scattering respectively. In the following we will 
assume that the 1DEG is sufficiently incommensurate so that Umklapp 
scattering may be neglected. Consequently the charge sector remains gapless 
and continues to be described by a free bosonic theory [the charge part of 
Eq. (\ref{diagh})]. Including a backward scattering term $g_1\sum_\eta
\psi_{\eta,1}^{\dagger}\psi_{-\eta,-1}^{\dagger}\psi_{\eta,-1}\psi_{-\eta,1}$
results, after bosonization, in a spin Hamiltonian density of the 
sine-Gordon type
\begin{equation}
\label{sghamiltonian}
{\cal H}_s=\frac{v_s}{2}\left[K_s(\partial_x\theta_s)^2+
\frac{(\partial_x\phi_s)^2}{K_s}\right]+\frac{2g_1}{(2\pi a)^2}
\cos(\sqrt{8\pi}\phi_s) \; .   
\end{equation}

The $g_1$ perturbation is relevant for $K_s<1$, in which case a spin gap is 
dynamically generated according to the scaling relation \cite{Gogolin98} 
$\Delta_s\sim(v_s/a)[g_1/2\pi^2 v_s]^{1/(2-2K_s)}$ and the excitations 
are massive spin solitons. In the spin gapped phase the problem is most 
simply treated in terms of spin fermion fields and their mode 
decomposition $C_\pm(\theta)$ (obeying the usual fermionic 
anti-commutation relations) in momentum space
\end{multicols}
\widetext

\noindent
\setlength{\unitlength}{1in}
\begin{picture}(3.375,0)
  \put(0,0){\line(1,0){3.375}}
  \put(3.375,0){\line(0,1){0.08}}
\end{picture}

\begin{eqnarray}
\label{spinfields}
\nonumber
\Psi_\eta&=&F_\eta\exp[-i\sqrt{\pi/2}(\theta_s-2\eta\phi_s)] \\
&=&\sqrt{\frac{\Delta_s}{4\pi v_s}}\int_{-\infty}^{\infty}d\theta 
e^{\eta\theta/2}\left[C_+(\theta)e^{i(x/\xi_s)\sinh\theta}-
\eta C_-^{\dagger}(\theta)e^{-i(x/\xi_s)\sinh\theta}\right] \; ,
\end{eqnarray}

\hfill
\begin{picture}(3.375,0)
  \put(0,0){\line(1,0){3.375}}
  \put(0,0){\line(0,-1){0.08}}
\end{picture}

\begin{multicols}{2}
\noindent
where we introduce the spin correlation length $\xi_s=v_s/\Delta_s$ 
and use the rapidity representation $k=\sinh\theta/\xi_s$.

For $K_s=1/2$, which is known as the free fermion or Luther-Emery 
point\cite{Luther74b}, the refermionized Hamiltonian is non-interacting 
and massive with a gap 
$\Delta_s=g_1/2\pi a$
\begin{eqnarray}
\label{refh}
\nonumber
H_s&=&\int dx \sum_{\eta=\pm 1} \left[-i v_s \eta\Psi_\eta^{\dagger}
\partial_x\Psi_\eta+\Delta_s\Psi_\eta^{\dagger}\Psi_{-\eta}\right] \\
\nonumber
&=&\int_{-\infty}^{\infty}d\theta E_s(\theta)\left[
C_+^{\dagger}(\theta)C_+(\theta)+C_-^{\dagger}(\theta)C_-(\theta)-1 
\right] \; , \\
\end{eqnarray}
where the spin excitation spectrum is 
\begin{equation}
\label{spinexcspec}
E_s(\theta)\equiv\Delta_s\cosh\theta=\sqrt{\Delta_s^2+(v_s k)^2}
\equiv E_s(k) \; .
\end{equation} 

In the following we will concentrate on computing the 
correlation functions at the Luther-Emery point. We will comment briefly 
on the effects of deviations from this point. 

\subsection{The spin part of the spectral functions}

In contrast to the Tomonaga-Luttinger model the calculation of the spin 
part of most spectral functions for the Luther-Emery liquid is no longer 
trivial. The difficulty lies in the fact that generically the refermionized 
form for the spectral functions involves highly non-local operators.
We start by evaluating the spin contribution to the transverse spin form 
factor, which fortunately has a simple representation in terms of the 
pseudo-fermions. We then consider the single hole spectral function that 
belongs to the wider class of functions which do not admit such a simple 
form. Interestingly, it is still possible to obtain an exact expression 
for this function too at the Luther-Emery point. This exact result extends
and corrects earlier work of Voit\cite{Voit98} and Wiegmann\cite{Wiegmann99}.

In computing the various spectral properties of the system we can distinguish
between two temperature regions. At temperatures large compared to $\Delta_s$
the spin gap can be ignored, and the results for the Tomonaga-Luttinger liquid
apply. If the temperature is small compared to the spin gap we can 
evaluate the spin contribution to the correlation functions in the zero 
temperature limit while introducing errors of order $\exp(-\Delta_s/T)$. 
Henceforth we will be concerned with temperatures in the second region only.

\subsubsection{The spin part of ${\cal S}(k,\omega)$}

The spin piece of the transverse spin correlation function has the 
following simple form in terms of the spin fermion fields
\begin{equation}
\label{spinexp}
\tilde{\cal S}_{s}(x,t) = \langle\Psi_1^{\dagger}(x,t)\Psi_{-1}^
{\dagger}(x,t)\Psi_{-1}(0,0)\Psi_1(0,0)\rangle \; .
\end{equation}
Since the theory reduces, at the Luther-Emery point, to a theory of free 
massive fermions, the corresponding spectral function can be readily computed 
with the result, for $T=0$,  
\begin{equation}
\label{eq:Sspngap}
{\cal S}_s(k,\omega)=\frac{\omega^2-4E_s^2(k/2)}{4v_s^2|q_1 E_s(q_2)
-q_2 E_s(q_1)|}\Theta[-\omega-2E_s(k/2)] \; ,
\end{equation}
where the spin excitation spectrum $E_s(k)$ is given by Eq.(\ref{spinexcspec})
and $q_{1,2}$ are the solutions to the quadratic equation  
$\omega+E_s(q)+E_s(k-q)=0$, that is
\begin{equation}
\label{q1q2s}
q_{1,2}=\frac{k}{2}\pm\frac{\omega}{2v_s}\sqrt{1+\frac{4\Delta_s^2}
{v_s^2 k^2-\omega^2}} \; .
\end{equation}

\subsubsection{The spin part of $G^<(k,\omega)$}

The refermionized form of the single hole Green function (below we consider
the case $\eta=-1$ and $\sigma=1$)
\begin{equation}
\label{refg}
\tilde G_{s}(x,t)=\langle U_{\frac{1}{4}}^{\dagger}(x,t)
\Psi_{-1}^{\dagger}(x,t)\Psi_{-1}(0,0)U_{\frac{1}{4}}(0,0)\rangle \; ,
\end{equation}
involves the non-local vertex operators
\begin{equation}
\label{vertex}
U_\alpha(x)=\exp[i\sqrt{8\pi}\,\alpha\phi_s(x)] \; ,
\end{equation}
with
\begin{equation}
\label{refphis}
\phi_s(x)=\sqrt{\frac{\pi}{2}}\sum_{\eta=\pm 1}\int^x dy 
\Psi_\eta^{\dagger}(y)\Psi_\eta(y) \; .
\end{equation}
From Eqs. (\ref{spinfields}), (\ref{vertex}) and (\ref{refphis}) it is evident 
that $\Psi_\eta$ and $U_\alpha$ create and destroy a single spin soliton 
or an integer number of soliton pairs, respectively. Therefor, the 
Green function consists of a coherent one spin soliton piece and an 
incoherent multi-soliton piece\cite{Carlson00}
\begin{equation}
\label{eq:Gs}
G_{s}(k,\omega)=  Z_s(k) \delta[\omega+E_{s}(k)]
+G_{s}^{(multi)}(k,\omega) \; ,
\end{equation}
where the multi-soliton piece is proportional (at $T=0$) to 
$\Theta[-\omega-3E_{s}(k/3)]$. Deviations from the Luther-Emery point  
in the case $K_s<1/2$ will result in the formation of a
spin soliton-antisoliton bound state, a ``breather'', 
which can shift the threshold energy for
multi-soliton excitations somewhat.

At the Luther-Emery point the form factors of the vertex operators, 
{\it i.e.} their matrix elements between the vacuum and various 
multi-soliton states, are known exactly. This fact enables us to 
obtain exact results for the different parts of the spectral function. 
The details of the calculation are presented in Appendix B. 
For the spectral weight of the coherent piece we find
\begin{equation}
\label{zs}
Z_s(k)=\frac{8c^2}{\pi}\left(\frac{2\xi_s}{a}\right)^{\frac{3}{8}}
\left[1-\frac{v_s k}{E_s(k)}\right] \; ,
\end{equation}
where $c=0.101$. We would like to note here that a 
simple scaling argument\cite{Carlson00} 
allows us to obtain the dependence of $Z_s$ on 
$\xi_s$ for arbitrary $K_s<1$. It follows from the observation that the 
sine-Gordon theory is asymptotically free and hence the dependence of 
$G_s$ on the short distance cutoff $a$ is unaffected 
by the opening of a spin gap. Since in the absence of a gap $G_s$ is 
proportional to $a^{2\gamma_s-1/2}$ 
it is a matter of dimensional analysis to see that 
\begin{equation}
\label{zsgeneral}
Z_s(k)=(\xi_s/a)^{\frac{1}{2}-2\gamma_s}f_s(k\xi_s) \;,
\end{equation}
where $f_s$ is an undetermined scaling function. 

The incoherent piece of $G_s^<(k,\omega)$ consists of contributions from 
processes involving intermediate states containing $N=2n+1$ spin solitons, 
$n=1,2,3,\cdots$. Each such contribution starts at an energy threshold 
$-N E_s(k/N)$ 
\begin{equation}
\label{gmultistruc}
G_s^{(multi)}(k,\omega)=\!\!\!\!\sum_{N=3,5,\cdots}
\!\!\!\!G_s^{(N\,{\rm sol})}(k,\omega) 
\,\Theta\!\left[-\omega-NE_s\!\!\left(\frac{k}{N}\right)\!\right] \,\! .
\end{equation}
At the vicinity of the threshold 
{\it i.e.} for $k\xi_s\ll 1$ and $|\omega+N E_s(k/N)|/\Delta_s\ll 1$ the 
behavior of $G_s^{(N\;{\rm sol})}$ is given 
by $[\omega+N E_s(k/N)]^{n^2+n-1}$. 
In particular the 3 soliton part reads then
\begin{equation}
\label{3sol}
G_s^{(3\;{\rm sol})}(k,\omega)=\frac{8c^2}{\sqrt{3}\pi^2}\left(\frac{2\xi_s}
{a}\right)^{\frac{3}{8}}\left[\frac{-\omega-3 E_s(k/3)}
{\Delta_s^2}\right] \; .
\end{equation}
As discussed  above, for large $|\omega|$ the incoherent piece should 
asymptotically approach the Tomonaga-Luttinger result 
$\omega^{2\gamma_s-3/2}$.

\subsection{The single hole spectral function}
Once the spin part is calculated there still remains the task of 
convolving it with the corresponding charge part in order to 
obtain an expression for the full spectral function. Analytically this 
is difficult, and like in the case of the Tomonaga-Luttinger liquid, 
progress can be made only in special cases. Below we carry out the 
convolution for the single hole spectral function. We consider temperatures 
well below the spin gap scale and correspondingly use the above derived 
zero temperature results for the spin part. For the gapless charge degrees
of freedom we continue to use the finite temperature Tomonaga-Luttinger 
expressions.      

The spectral function can be written as a sum of two contributions 
$G^<=G_1+G_2$ coming from the convolution of the charge part with the 
coherent (single soliton) and incoherent (multi-soliton) 
pieces of $G_s^<$. The frequency integral in 
$G_1$ is readily evaluated with the result
\begin{eqnarray}
\label{g1general}
\nonumber
G_1&&(k,\omega;T)=\frac{c^2}{\pi^3}\frac{\lambda_T^2}{v_c}\left(\frac{a}
{\lambda_T}\right)^{2\gamma_c+\frac{1}{2}}\left(\frac{2\xi_s}{a}\right)
^{\frac{3}{8}} \\
\nonumber 
\times && \int dq \left[1-\frac{v_s(k-q)}{E_s(k-q)}\right] 
h_{\gamma_c+\frac{1}{2}}\left[\frac{\omega+E_s(k-q)+v_c q}{2\pi T}\right] \\
\times && h_{\gamma_c}\left[\frac{\omega+E_s(k-q)-v_c q}{2\pi T}\right] \; , 
\end{eqnarray}   
where here $\lambda_T=v_c/\pi T$. 

In Fig. 2 we present representative contributions of the single spin 
soliton piece, $G_1$, to the MDCs and EDCs of a Luther-Emery liquid with 
various values of the charge exponent $\gamma_c$. Using some of the results 
derived below, we also indicate the 
asymptotic behavior of the three spin soliton contribution, 
$G_2^{(3\;{\rm sol})}$, to the EDCs in the vicinity of its zero 
temperature threshold $\omega=-3\Delta_s$.    

\begin{figure}
\narrowtext
\setlength{\unitlength}{1in}
\begin{picture}(3.2,4.2)(0,-1.2)
\put(-0.15,-1){\psfig{figure=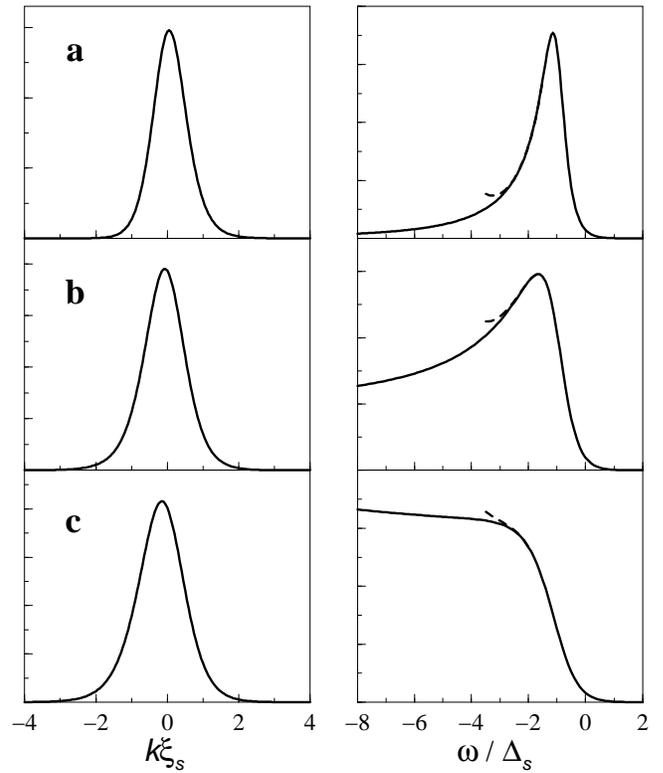,width=3.4in}}
\end{picture}
\caption
{The single spin soliton contribution, $G_1$, to the MDCs 
at $\omega=0$ (left) and EDCs at $k=0$ (right), of a Luther-Emery 
liquid $(K_s=1/2)$, with $v_c/v_s=3$, $\Delta_s/T=3$ and a) $\gamma_c=0$, 
b) $\gamma_c=0.2$, and c) $\gamma_c=0.4$. The asymptotic contribution 
of the three spin soliton piece, $G_2^{(3\;{\rm sol})}$, to the EDCs, near 
its zero temperature threshold $\omega=-3\Delta_s$, is indicated by the 
dashed lines. This contribution was calculated using Eq. (\ref{g1g0general}) 
in the case $\gamma_c=0$. For the EDCs with $\gamma_c=0.2,0.4$ we used the 
asymptotic result for the case $v_s/v_c\rightarrow 0$, Eq. 
(\ref{g23solgeneral}). Other multi-soliton processes start to 
contribute near $\omega/\Delta_s=5,7,\cdots$, as discussed in the text.}
\label{fig:2}
\end{figure}

We now restrict the discussion to the 
following special circumstances:   

\subsubsection{The case $v_s/v_c\rightarrow 0$}

In this case the remaining integral in (\ref{g1general}) is straightforward 
since the $q$ dependence of the spin part and $E_s(k-q)$ disappears and 
we are left with 
\begin{eqnarray}
\label{g1r0}
\nonumber
G_1(k,\omega;T)&=&\frac{4c^2}{\pi^2}\frac{a}{v_c}\left(\frac{a}{\lambda_T}
\right)^{2\gamma_c-\frac{1}{2}}\left(\frac{2\xi_s}{a}\right)^{\frac{3}{8}} \\
&\times&\left[1-\frac{v_s k}{E_s(k)}\right]h_{2\gamma_c+\frac{1}{2}}
\left[\frac{\omega+E_s(k)}{\pi T}\right] \; ,
\end{eqnarray}
which in the limit of zero temperature reduces to 
\begin{eqnarray}
\label{g1r0t0}
\nonumber
G_1(k,\omega;0)&=&\frac{8c^2}{\pi}\frac{1}{\Gamma(2\gamma_c+1/2)}
\left(\frac{a}{v_c}\right)^{2\gamma_c+\frac{1}{2}}
\left(\frac{2\xi_s}{a}\right)^{\frac{3}{8}} \\
&\times&\left[-\omega-E_s(k)\right]^{2\gamma_c-\frac{1}{2}}
\Theta[-\omega-E_s(k)] \; .
\end{eqnarray}

For the 3-soliton contribution to $G_2$ we obtain, assuming $k\xi_s\ll 1$ and 
$|\omega+3E_s(k/3)|\ll\Delta_s$, 
\begin{eqnarray}
\label{g23solgeneral}
\nonumber
&&G_2^{(3\;{\rm sol})}(k,\omega;T)=\frac{4c^2}{\sqrt{3}\pi^4}
\frac{1}{\Delta_s T}
\left(\frac{a}{\lambda_T}\right)^{2\gamma_c+\frac{1}{2}}
\left(\frac{2\xi_s}{a}\right)^{\frac{3}{8}} \\
\nonumber
&&\times\int_{\omega+3E_s(k/3)}^{\infty} d\nu\, 
h_{2\gamma_c+\frac{1}{2}}\left(\frac{\nu}{\pi T}\right)\left[
\frac{\nu-\omega-3E_s(k/3)}{\Delta_s}\right] \;  , \\
\end{eqnarray}
which at $T=0$ equals
\begin{eqnarray}
\label{g23solt0}
\nonumber
G_2^{(3\;{\rm sol})}&&(k,\omega;0)=\!\frac{8c^2}{\sqrt{3}\pi^2}
\frac{1}{\Gamma(2\gamma_c+5/2)}\!\left(\frac{a}{v_c}\right)
^{2\gamma_c+\frac{1}{2}}\!\!\!\left(\frac{2\xi_s}{a}\right)^{\!\frac{3}{8}} \\
\nonumber
&&\times\frac{1}{\Delta_s^2} 
\left[-\omega-3E_s(k/3)\right]^{2\gamma_c+\frac{3}{2}}
\Theta[-\omega-3E_s(k/3)] \; . \\
\end{eqnarray}
For this case the $N=2n+1$ soliton contribution to $G_2$ is proportional, 
in the vicinity of its threshold, to $[-\omega-NE_s(k/N)]^
{2\gamma_c+n^2+n-1/2}\Theta[-\omega-NE_s(k/N)]$. 

\subsubsection{the case $\gamma_c=0$}
It is also possible to derive closed expressions for the spectral function 
when $\gamma_c=0$ but $r=v_s/v_c$ is arbitrary. 
\begin{eqnarray}
\label{g1g0general}
\nonumber
G_1(k,\omega\!&;&\!\!T)=\frac{4c^2}{\pi^2}\frac{1}{1-r}
\frac{\sqrt{a\lambda_T}}{v_c}
\left(\frac{2\xi_s}{a}\right)^{\frac{3}{8}} \\
\nonumber
&\times& h_{\frac{1}{2}}\left[\frac{\omega-r v_s k +\sqrt{(r\omega-v_s k)^2
+(1-r^2)\Delta_s^2}}{\pi(1-r^2)T}\right] \\
&\times&\frac{r\omega-v_s k+\sqrt{(r\omega-v_s k)^2+(1-r^2)\Delta_s^2}}
{\sqrt{(r\omega-v_s k)^2+(1-r^2)\Delta_s^2}} \; .
\end{eqnarray}
Using the asymptotic form (\ref{ztlimit}) of $h_{\gamma}(k)$ one can 
obtain the zero temperature limit of $G_1$. When $r<1$ we find
\begin{eqnarray}
\label{g1g0rgt0}
\nonumber
G_1(k,\omega\!&;&\!\!0)=\frac{8c^2}{\pi^{3/2}}\left(\frac{1+r}{1-r}\right)
^{\frac{1}{2}}\left(\frac{a}{v_c}\right)^{\frac{1}{2}}
\left(\frac{2\xi_s}{a}\right)^{\frac{3}{8}} \\
\nonumber
&\times&\frac{r\omega-v_s k+\sqrt{(r\omega-v_s k)^2+(1-r^2)\Delta_s^2}}
{\sqrt{(r\omega-v_s k)^2+(1-r^2)\Delta_s^2}} \\
\nonumber
&\times&\frac{\Theta[-\omega-E_s(k)]}{\sqrt{-\omega+r v_s k-
{\sqrt{(r\omega-v_s k)^2+(1-r^2)\Delta_s^2}}}} \; . \\ 
\end{eqnarray}
The kinematics in the case $r>1$ is more involved. In particular we find 
that for sufficiently large $r$ it is possible to distribute the total 
momentum between the spin soliton and the gapless charge modes in a way 
that gives a contribution at frequencies smaller than $E_s(k)$. (In this 
case the effect exists only for $k<0$. This is a special property 
of the point $\gamma_c=0$ which precludes mixing of left and right moving 
charge excitations.) 
\end{multicols}
\widetext

\noindent
\setlength{\unitlength}{1in}
\begin{picture}(3.375,0)
  \put(0,0){\line(1,0){3.375}}
  \put(3.375,0){\line(0,1){0.08}}
\end{picture}

\begin{eqnarray}
\label{g1g0rst0}
\nonumber
G_1(k,\omega;0)&=&\frac{8c^2}{\pi^{3/2}}\left(\frac{r+1}{r-1}\right)
^{\frac{1}{2}}\left(\frac{a}{v_c}\right)^{\frac{1}{2}}
\left(\frac{2\xi_s}{a}\right)^{\frac{3}{8}}\\
\nonumber
&\times&\sum_\sigma\frac{v_s k-r\omega-\sigma
\sqrt{(r\omega-v_s k)^2+(1-r^2)\Delta_s^2}}
{\sqrt{(r\omega-v_s k)^2+(1-r^2)\Delta_s^2}
\sqrt{\omega-r v_s k+\sigma\sqrt{(r\omega-v_s k)^2+(1-r^2)\Delta_s^2}}} \\
&&\left\{ 
\begin{array}{ll}
\sigma=1 & {\rm if}\;\; \omega<-E_s(k) \\
\sigma=-1,1 & {\rm if}\;\;  k\xi_s<-\frac{1}{\sqrt{r^2-1}}\;\; {\rm and} \;\;
-E_s(k)<\omega<\frac{v_s k-\sqrt{r^2-1}\Delta_s}{r} \\
\end{array}
\right. 
\; . 
\end{eqnarray}

\hfill
\begin{picture}(3.375,0)
  \put(0,0){\line(1,0){3.375}}
  \put(0,0){\line(0,-1){0.08}}
\end{picture}

\begin{multicols}{2}
\noindent

For $k\xi_s\ll 1$ and $|\omega+3E_s(k/3)|\ll\Delta_s$ the 3-soliton 
contribution to $G_2$ reads
\begin{eqnarray}
\label{g2g03sol}
\nonumber
G_2^{(3\;{\rm sol})}(k,\omega;T)&=&\frac{4c^2}{\sqrt{3}\pi^3}
\frac{\sqrt{a\lambda_T}}{\Delta_s}
\left(\frac{2\xi_s}{a}\right)^{\frac{3}{8}}\int dq\, h_{\frac{1}{2}}
(\lambda_T q) \\
\nonumber
&\times&\left[\frac{v_c q-\omega-3E_s[(k-q)/3]}{\Delta_s}\right] \\
&\times&\Theta\{v_c q-\omega-3E_s[(k-q)/3]\} \; . 
\end{eqnarray}
When $r<1$ the zero temperature limit of the above is readily evaluated 
giving
\begin{eqnarray}
\label{g23solt0rl}
\nonumber
G_2^{(3\;{\rm sol})}&&(k,\omega;0)=\frac{32c^2}{3^{3/2}\pi^{5/2}}
\left(\frac{a}{\Delta_s v_c}\right)^{\frac{1}{2}} 
\left(\frac{2\xi_s}{a}\right)^{\frac{3}{8}} \\
&&\times\left[-\frac{\omega+3E_s(k/3)}{\Delta_s}\right]^{\frac{3}{2}}
\Theta[-\omega-3E_s(k/3)] \; .
\end{eqnarray}
The zero temperature limit when $r>1$ is, once again, more complicated. 
However, it is possible to show that in the range of validity of Eq. 
(\ref{g2g03sol}), {\it i.e.} for $k\xi_s\ll 1$, and for $1<r\lesssim 3$ it 
coincides with the expression for the case $r<1$ Eq. (\ref{g23solt0rl}).
Deviations from this behavior may occur only if the velocity ratio is large 
and satisfies $r\gg 3$.

\acknowledgments{It is a pleasure to thank S.~Kivelson for many useful 
discussions and comments.}

\appendix

\section{The function $\lowercase{h}_\gamma(\lowercase{k})$}

The function $h_\gamma(k)$ is real 
\begin{eqnarray}
\label{hk}
\nonumber
h_\gamma(k)&=&\int_{-\infty}^{\infty}dx\,e^{ikx} h_{\gamma}(x) \\
&=& \lim_{a\rightarrow 0}2{\rm Re}\left\{\int_0^{\infty}dx\,\frac
{e^{ikx}}{[-i\sinh(x+ia)]^{\gamma}}\right\} \; .
\end{eqnarray}
Although the imaginary part of the integral diverges as $a^{1-\gamma}$ 
for $\gamma>1$, its real part, and hence also $h_\gamma(k)$, are analytic 
for all values of $\gamma$. Substituting $y=e^{-2x}$ we find
\begin{equation}
\label{hk1}
h_\gamma(k)=\lim_{a\rightarrow 0}{\rm Re}\left\{(2i)^\gamma 
\int_0^1\!dy\,y^{\frac{\gamma-ik}{2}-1}\left(1-e^{-2ia}y\right)
^{-\gamma}\right\} .
\end{equation}
The integral is analytic in the limit $a\rightarrow 0$ for $\gamma<1$. 
In this range the exponential factor in (\ref{hk1}) can be dropped and 
it reduces to the integral representation of the beta function \cite
{Gradshteyn94}. We can then use analytical continuation to obtain for 
all non-integer values of $\gamma$
\begin{equation}
\label{hk2}
h_\gamma(k)={\rm Re}\left[(2i)^\gamma B\left(\frac{\gamma-ik}{2},1-\gamma
\right)\right] \; .
\end{equation}

For the integers, $h_n(k)$ can be calculated as follows. First, we 
obtain using standard residue technique
\begin{eqnarray}
\label{h1h2}
h_1(k)&=&2\pi f_+(\pi k) \; , \\
h_2(k)&=&2\pi k f_-(\pi k) \; ,
\end{eqnarray}
where $f_\pm(k)=\left(e^k\pm 1\right)^{-1}$ are the fermionic and bosonic
occupation functions. We then integrate (\ref{hk}) by parts 
twice to find the recursion relation
\begin{equation}
\label{recursion}
h_{n+2}(k)=\frac{k^2+n^2}{n(n+1)}h_n(k) \; ,
\end{equation}
which implies
\begin{eqnarray}
\label{hn1hn2}
\nonumber
h_{2n+1}(k)&=&\frac{2\pi}{\Gamma(2n+1)}\prod_{m=0}^{n-1}\left[(1+2m)^2+k^2
\right]f_+(\pi k) \; , \\
h_{2n}(k)&=&\frac{2\pi}{\Gamma(2n)}k\prod_{m=1}^{n-1}\left[(2m)^2+k^2
\right]f_-(\pi k) \; . 
\end{eqnarray}
We also note that 
\begin{equation}
\label{h0k}
h_0(k)=2\pi\delta(k) \; .
\end{equation}

Finally, the asymptotic behavior of $h_\gamma(k)$ for large $|k|$ is 
easily evaluated with the result
\begin{equation}
\label{ztlimit}
h_\gamma(|k|\rightarrow\infty)=\frac{2\pi}{\Gamma(\gamma)}\Theta(-k)
(-k)^{\gamma-1} \; .
\end{equation}
Since the spectral functions have a scaling form in the variables $\omega/T$ 
and $v_\alpha k/T$ Eq. (\ref{ztlimit}) also determines the zero temperature 
limit of these functions.

\section{Calculating $G_{\lowercase{s}}^<(\lowercase{k},\omega)$ 
at the Luther-Emery point}

Inserting the resolution of the identity into Eq. (\ref{refg}) one obtains
\begin{eqnarray}
\label{relrefg}
\nonumber
\tilde G_s^<(x,t)=\sum_{N,N'}\langle 0|U_{\frac{1}{4}}^\dagger(x,t)|N&&\rangle 
\langle N|\Psi_{-1}^\dagger(x,t)\Psi_{-1}(0,0)|N' \rangle \\
&&\times\langle N'|U_{\frac{1}{4}}(0,0)|0\rangle \; .
\end{eqnarray}

The matrix elements of the vertex operators appearing in (\ref{relrefg}) are 
known as the form factors of these operators. They have been derived, at the 
Luther-Emery point (see also Ref. \ref{Lukref}), by a variety of ways. They 
were first obtained by Schroer and Truong\cite{Schroer78} who normal ordered 
the vertex operator with respect to the spin fermions. Smirnov\cite{Smirnov92}
derived them for $\alpha=1/2$ using bootstrap axioms. Most recently they 
were calculated using monodromy relations by Bernard and LeClair
\cite{Bernard94}. Since the fields $U_\alpha(0,0)$ are neutral with respect 
to the topological $U(1)$ charge of the solitons $(\pm)$ the form factors 
are non-vanishing only for $U(1)$ neutral states
\end{multicols}
\renewcommand{\theequation}{B\arabic{equation}}
\widetext

\noindent
\setlength{\unitlength}{1in}
\begin{picture}(3.375,0)
  \put(0,0){\line(1,0){3.375}}
  \put(3.375,0){\line(0,1){0.08}}
\end{picture}

\begin{eqnarray}
\label{formfact}
\nonumber
\langle 0|U_\alpha&&(x,t)C_+^\dagger(\theta_{2n})\cdots 
C_+^\dagger(\theta_{n+1})C_-^\dagger(\theta_{n})\cdots
C_-^\dagger(\theta_{1})|0\rangle = 
V_\alpha\,(-1)^{n(n-1)/2}\left(\frac{\sin \pi\alpha}
{2\pi i}\right)^n \\
\times\,&&\exp\left[i\sum_{k=1}^{2n}\left(\frac{x}{\xi_s}\sinh\theta_k-
\Delta_s t\cosh\theta_k\right)+
\alpha\sum_{k=1}^n(\theta_{n+k}-\theta_k)\right] 
\frac{\prod_{1\leq k<j\leq n}\sinh\left(\frac{\theta_k-\theta_j}{2}\right)
\sinh\left(\frac{\theta_{n+k}-\theta_{n+j}}{2}\right)}
{\prod_{1\leq k,j\leq n}\cosh\left(\frac{\theta_{n+k}-\theta_j}{2}\right)} \; ,
\end{eqnarray}

\hfill
\begin{picture}(3.375,0)
  \put(0,0){\line(1,0){3.375}}
  \put(0,0){\line(0,-1){0.08}}
\end{picture}

\begin{multicols}{2}
\noindent
where $V_\alpha$, the vacuum expectation value of the vertex operators, 
is given by 
\cite{Lukyanov97b}
\begin{equation}
\label{vev}
V_\alpha\equiv\langle 0|U_\alpha(0,0)|0\rangle=
c(\alpha)\left(\frac{2\xi_s}{a}\right)^{-\alpha^2} \; ,
\end{equation}
with 
\begin{equation}
\label{cconst}
c(\alpha)=\exp\left\{\int_0^\infty \frac{dt}{t}\left[\frac{\sinh^2(\alpha t)}
{\sinh^2 t}-\alpha^2 e^{-2t}\right]\right\} \; .
\end{equation} 

Furthermore, since $\Psi_\eta$ creates and destroys a single soliton 
the state $|N'\rangle$ can differ from $|N\rangle$ by $0,\pm2$ solitons 
only. It is also easy to check that $G_s^<(k,\omega)$ is real. Using 
Eqs. (\ref{spinfields}) and (\ref{formfact}) one obtains that each 
term in the Fourier transform of Eq. (\ref{relrefg}) is proportional to 
$i^{(N+N')/2}$ times a real expression. Thus only terms with $(N+N')/2$ 
an even integer should be considered. Combining these two observations 
we conclude that the contribution to $G_s^<(k,\omega)$ comes solely from 
the terms $N=N'$ (the amplitudes for the cases $N=N'\pm 2$ are 
finite but cancel each other). 

The coherent piece of the spectral function is due to the terms $N=0$ and 
$N=2$. The evaluation of the first is straightforward with the result
\[
\pi V^2\left[1-\frac{v_s k}{E_s(k)}\right]\delta[\omega+E_s(k)] \; , 
\]
where $V=V_{\frac{1}{4}}=c(\frac{1}{4})(2\xi_s/a)^{-1/16}$. The cutoff 
dependence of the above result is in conflict with the general scaling 
argument that was given following Eq. (\ref{zs}). It represents corrections 
to scaling coming from irrelevant operators. The leading scaling behavior 
is recovered by considering the contribution of the $N=2$ term to the 
coherent piece
\[
\frac{V^2}{32\pi^3\xi_s}\left[\int d\beta\frac{e^{-\frac{3}{4}\beta}}
{\cosh\frac{\beta}{2}}\right]^2 \!\int d\theta\,e^
{-\theta+i[(x/\xi_s)\sinh\theta-
\Delta_s t\cosh\theta]} \; .
\]
The $\beta$ integral is divergent and a lower cutoff $-\ln(2\xi_s/a)$ 
(corresponding to a cutoff $k=-1/a$) should be introduced. We then find 
for the leading contribution to the coherent piece
\begin{eqnarray}
\label{coherxt}
\nonumber
\tilde G_s^{(coher)}(x,t)=&&\frac{4c^2}{i\pi^3\xi_s}
\left(\frac{2\xi_s}{a}\right)^{\frac{3}{8}}\frac{x-v_s t}
{\sqrt{x^2-(v_s t)^2+i\epsilon t}} \\
&&\times K_1\left[\frac{\sqrt{x^2-(v_s t)^2+i\epsilon t}}{\xi_s}\right] \; ,
\end{eqnarray}
where $K_1(x)$ is a modified Bessel function. Fourier transforming it one 
obtains $Z_s(k) \delta[\omega+E_{s}(k)]$ with the spectral weight $Z_s(k)$ 
given by Eq. (\ref{zs}).

The $2n+1$ soliton contribution to the incoherent piece of the spectral 
function comes from the terms $N=2n$ and $N=2n+2$. As is the case with 
the coherent piece the former contains corrections to scaling while the 
latter is responsible for the leading behavior which is proportional to 
\end{multicols}
\renewcommand{\theequation}{B\arabic{equation}}
\widetext

\noindent
\setlength{\unitlength}{1in}
\begin{picture}(3.375,0)
  \put(0,0){\line(1,0){3.375}}
  \put(3.375,0){\line(0,1){0.08}}
\end{picture}

\begin{eqnarray}
\nonumber
V^2 \int d\theta_1\cdots d\theta_{2n+1}&& 
\left[\int d\beta \,e^{-\frac{3}{4}\beta}
\frac{\prod_{n+2\leq j\leq 2n+1}\sinh\left(\frac{\theta_j-\beta}{2}\right)}
{\prod_{1\leq j\leq n+1}\cosh\left(\frac{\theta_j-\beta}{2}\right)}\right]^2
\delta\left(k\xi_s-\sum_{i=1}^{2n+1}\sinh\theta_i\right)
\delta\left(\frac{\omega}{\Delta_s}+\sum_{i=1}^{2n+1}\cosh\theta_i\right) \\
\nonumber
\times&&\exp\left(\frac{1}{2}
\sum_{j=1}^{n+1}\theta_j-\frac{1}{2}\sum_{j=n+2}^{2n+1}\theta_j\right) 
\frac{\prod_{1\leq k<j\leq n+1}\sinh^2\left(\frac{\theta_k-\theta_j}{2}
\right)\prod_{n+2\leq k<j\leq 2n+1}\sinh^2\left(\frac{\theta_k-\theta_j}{2}
\right)}{\prod_{n+2\leq k\leq 2n+1\, , \, 1\leq j\leq n+1}
\cosh^2\left(\frac{\theta_k-\theta_j}{2}\right)} \; .
\end{eqnarray}

\hfill
\begin{picture}(3.375,0)
  \put(0,0){\line(1,0){3.375}}
  \put(0,0){\line(0,-1){0.08}}
\end{picture}

\begin{multicols}{2}
\noindent
The exact evaluation of the above integrals is difficult but their behavior 
near the threshold $(2n+1)E_s[k/(2n+1)]$ may be extracted in the following way.
Writing $\omega=-(2n+1)E_s[k/(2n+1)]-\Delta\omega$ one finds that for 
$k\xi_s\ll 1$ and $\Delta\omega\ll\Delta_s$ the second $\delta$-function 
restricts the integration region over $\theta_i$ to a small ball near the 
origin. It is then legitimate to expand the integrand to lowest order in 
these variables. The integral over $\beta$ gives (together with the factor 
$V^2$) the overall cutoff dependence $(2\xi_s/a)^{3/8}$ in accord with the 
general scaling argument. By changing variables
to spherical coordinates it is easy to check that the remaining integral 
over $\theta_i$ is proportional to $\Delta\omega^{n^2+n-1}\Theta(\Delta
\omega)$. A detailed evaluation of the 3 solitons case results in 
Eq. (\ref{3sol}).


\end{multicols}

\end{document}